\documentclass[10pt,conf]{IEEEtran}
\usepackage{geometry}
\usepackage{bbold}
\usepackage{float}
\usepackage{subcaption}%
\usepackage[cmex10]{amsmath}
\usepackage{amssymb}
\usepackage{verbatim}
\usepackage{array}
\usepackage{latexsym}
\usepackage{color}
\usepackage{tabularx} 
\usepackage{textgreek}
\usepackage{subscript}
\usepackage{rotating}
\usepackage{booktabs,multirow}
\usepackage{mdframed}	
\usepackage{tabularx}
\usepackage{amsmath}
\usepackage{makecell}
\usepackage{tabto}

\usepackage{mathrsfs}
\usepackage{url}
\usepackage{cite}
\usepackage{listings}
\usepackage{array}
\usepackage[table]{xcolor}

\newcommand{\nref}[1]{\textcolor{red}{[{\bf NEED CITATION}]}}
\begin{document}
\setlength{\columnsep}{0.2 in}
\def\BibTeX{{\rm B\kern-.05em{\sc i\kern-.025em b}\kern-.08em T\kern-.1667em\lower.7ex\hbox{E}\kern-.125emX}}

\title{On the Interplay Between Network Metrics and Performance of Mobile Edge Offloading}

\author{
	~Parisa Fard Moshiri, Murat~Simsek,~Burak Kantarci
	\thanks{
	The authors are with the School of Electrical Engineering and 
        Computer Science at the University of Ottawa, Ottawa, ON, K1N 6N5, Canada.
		E-mail: \{ parisa.fard.moshiri,murat.simsek,burak.kantarci\}@uottawa.ca }

}

\maketitle
\thispagestyle{empty}
\pagestyle{empty}
\begin{abstract}

Multi-Access Edge Computing (MEC) emerged as a viable computing allocation method that facilitates offloading tasks to edge servers for efficient processing. The integration of MEC with 5G, referred to as 5G-MEC, provides real-time processing and data-driven decision-making in close proximity to the user.
The 5G-MEC has gained significant recognition in task offloading as an essential tool for applications that require low delay. Nevertheless, few studies consider the dropped task ratio metric. Disregarding this metric might possibly undermine system efficiency. In this paper, the dropped task ratio and delay has been minimized in a realistic 5G-MEC task offloading scenario implemented in NS3. We utilize Mixed Integer Linear Programming (MILP) and Genetic Algorithm (GA) to optimize delay and dropped task ratio. We examined the effect of the number of tasks and users on the dropped task ratio and delay. Compared to two traditional offloading schemes, First Come First Serve (FCFS) and Shortest Task First (STF), our proposed method effectively works in 5G-MEC task offloading scenario.  For MILP, the dropped task ratio and delay has been minimized by 20\% and  2ms compared to GA. 
\end{abstract}

\begin{IEEEkeywords}
Mobile Edge Computing, Task Offloading, 5G, Optimization 
\end{IEEEkeywords}

%
\IEEEpeerreviewmaketitle
\vspace{-2mm}
\section{Introduction}
%
%
%
%
Despite ongoing improvements in the Central Processor Unit (CPU), battery capacity, and other software and hardware components of smartphones, their capabilities remain constrained by physical design limitations. Consequently, they cannot efficiently run applications requiring extensive computation within a limited timeframe. Accordingly, many researchers have turned to mobile edge computing (MEC) to address the challenges \cite{haibeh.2022}.

The introduction of 5G technology has brought out  several improvements such as high-speed data transfer and ultra-low latency. These enhancements have the ability to address the limitations noticed in current mobile terminal devices when it comes to computation-intensive tasks. These limitations include processing bottlenecks, latency issues, and bandwidth restrictions \cite{kumaran.2021}. The combination of 5G Enhanced Mobile Broadband (eMBB) and Multi-access Edge Computing (MEC) offers significant advantages to real-time video and image processing applications, such as distracted driving recognition, pedestrian detection, and traffic sign detection. The eMBB is primarily concerned with achieving high data rates, making it particularly suitable for handling heavy data tasks. Moreover, URLLC (Ultra-Reliable and Low Latency Communications) guarantees a minimal latency between the acquisition of data and its processing, which is crucial for  detection in real-time scenarios \cite{nencioni2023}. Meanwhile, mMTC (massive Machine-Type Communications) enhances scalability by facilitating communication between a wide range of devices. Together, while MEC provides the computational power, 5G's characteristics, particularly eMBB and URLLC, ensure rapid and reliable data transmission, allowing for timely and accurate alerts\cite{Cruz2022}. 

 Computational task offloading is widely used approach in MEC system implementation since it reduces latency, enhances bandwidth efficiency, supports real-time processing, and optimizes resource use, benefiting mobile applications \cite{hua2023edge}. Computational offloading involves transferring processing tasks from user devices to MEC, servers, or cloud data centers \cite{haibeh.2022}. The offloading strategy is usually determined by the computational capability of mobile devices and the resources available in MEC and cloud servers \cite{sharma.2022}.

The potential for task offloading in MEC has recently grown quite dense due to the fact that a wide range of computational tasks can benefit from being processed closer to the edge \cite{feng.2021}. MEC has a substantial application in deep neural network-based video and image processing, which involves high computational cost \cite{feng.2021}. Mobile devices may struggle to provide real-time analysis, high data throughput, and low latency due to restricted resources. MEC provides an approach that combines proximity with powerful computing capability \cite{truong.2020}. When users offload computing tasks to the MEC server, several overhead factors arise, including task offloading delay, computation delay, and task completion rate \cite{feng.2021,zhu.2021}. During offloading, tasks are held in a memory queue, resulting in a delay in their execution and also dropping some of the tasks when they don’t meet the deadline\cite{Cruz2022}. Offloading strategies and resource allocation are critical for achieving efficiency in the offloading process, including optimal resource usage, minimized latency, energy conservation, optimal network utilization, and cost-effectiveness \cite{liao.2021}.

To the best of our knowledge, the idea of dropped task ratio, resulting from resource limitations, has been taken into account by few studies as a metric for optimization along with other metrics like delay, based on a comprehensive review of the published works on the topic. By considering this metric in the optimization process, more efficient decisions can be made while taking into account the system's limitations. This ensures that tasks are not just processed quickly (low delay) but are also less likely to be dropped due to insufficient resources or deadlines. Thus, when aiming to achieve the most efficient outcomes in terms of reducing delay, it is imperative to take into account the dropped task ratio. 

Managing the sequential execution of tasks is a considerable challenge when confronted with a substantial volume of tasks that are susceptible to being omitted or experiencing substantial delays. Various optimization strategies can effectively reduce this cost \cite{truong.2020, hsu.2022, haibeh.2022,kumaran.2021}. These techniques encompass traditional methods such as linear programming and dynamic programming, as well as contemporary heuristic algorithms like Genetic Algorithms (GA)\cite{zhu.2021, liao.2021}, and machine learning-based approaches \cite{truong.2020,liu.2019, hsu.2022,kumaran.2021}. The primary objective in these works often revolves around minimizing delay, latency, or energy. However, the dropped task ratio along with other metrics often goes unaddressed. In this study, we utilize optimization techniques to minimize not only computing delay but also the often-overlooked metric of dropped task ratio. This methodology provides a more comprehensive evaluation of system efficacy, achieving an appropriate balance between speed and reliability. The main contributions of this paper are summarized as follows:
\begin{enumerate}
        \item [1)] This model focuses on the offloading of the computationally intensive tasks of distracted driver recognition. We investigate the impact of networking parameters, including the number of users and  tasks on delay and dropped task ratio. Moreover, the effect of these parameters on different scheduling algorithms has been analyzed.
        \item [2)]  An optimization problem is set up in line with our scenario, targeting the minimization of delay and the dropped task ratio. We employ two optimization techniques, including Mixed Integer Linear Programming (MILP) and GA. We compare the optimization results with two scheduling algorithms, First Come First Serve (FCFS) and Shortest Task First (STF).   
    \end{enumerate}
The remainder of the paper is organized as follows. Section II
presents the literature review. Section III discusses the system
model and motivates the problem. Section IV presents and analyzes performance. We conclude in Section V.

\begin{figure*}[!hbt]
        \centering
        \includegraphics[width = 0.8\textwidth, trim=0cm 4cm 0cm 3cm,clip]{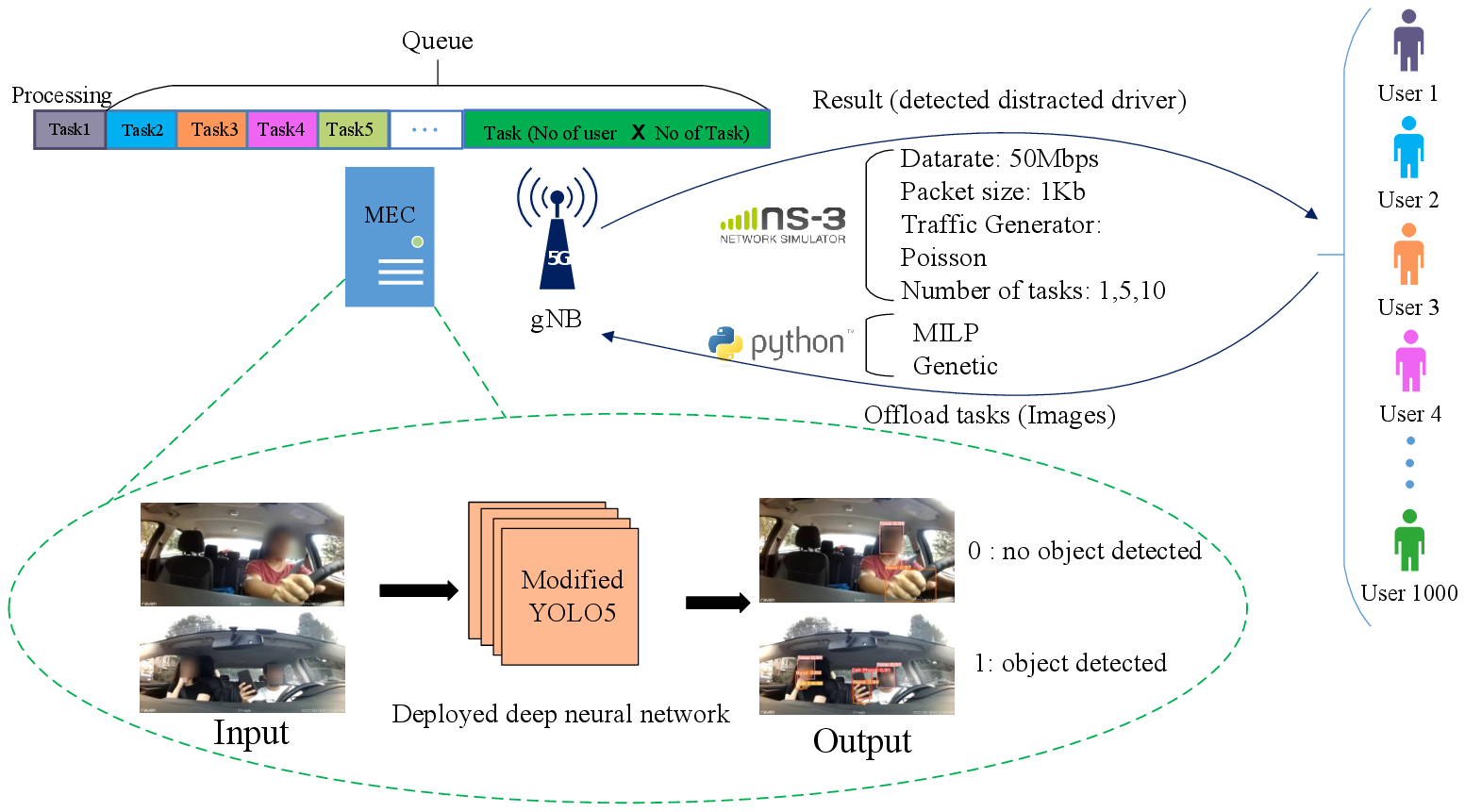}
        \caption{ Proposed methodology}
        \label{fig:methodology} \vspace{-2mm}
\end{figure*}
\vspace{-2mm}
\section{Related Work}
The existing literature on computational offloading mostly focuses on either the decision-making challenge of offloading tasks for multiple users or the allocation of wireless and computer resources. A DRL-based optimization for blockchain-enabled IoT resource allocation is proposed by Liu et al. \cite{liu.2019}. Wang et al. \cite{wang.2019} present IoT resource allocation using distributed-Q-learning. These researchers don't thoroughly examine the implications of offloading decision-making and resource allocation on computational offloading performance, resulting in challenges in achieving optimal efficiency, a balance between resource utilization, latency, energy, and network usage.

 Zhu et al.\cite{zhu.2021} analyze the effect of offloading decisions, uplink power allocation related, and computing resource allocation on system performance. The optimization function, solved by improved GA, is defined as the weighted sum of task execution latency and energy usage. Hsu et al. \cite{hsu.2022} study computation offloading in a heterogeneously loaded MEC-enabled Internet of Things (IoT) network. This work models the compute offloading as a Markov decision problem and develops a Dueling-Deep Q-learning Network (DQN) based energy and latency-efficient technique.
 
In some cases, researchers consider Multiple servers for MEC. However, the dense deployment of Base Stations (BS) might lead to several BS covering the end device, creating issues for offloading decisions. These decisions include determining where to offload computational workloads for low latency and energy cost \cite{liao.2021}. The Multi-User Multi Server edge computing challenge in ultra-dense cellular networks is addressed in \cite{liao.2021}. The mobile users are assigned to one BS based on physical distance and workload. The original problem is broken into parallel multi-user-to-one-server offloading choice subproblems after grouping. The average latency and energy are minimized by using a binary-coded GA. They analyze the effect of CPU frequency and the number of users on system performance. However, they do not consider the effect of the number of tasks assigned to each user on system performance. Their model is suitable in cases when we don't have many users or large amount of tasks. In another work, by Yuan et al.\cite{yuan.2022}, DQN and Simulated Annealing (SA) are employed to efficiently offload computing tasks. The SA is also integrated into the DQN Algorithm, and this algorithm effectively reduces the delay
and energy consumption of task processing.

In multi-server scenarios, load distribution in resource allocation is crucial. MILP can be used to determine the optimal allocation of tasks among the available resources. It achieves an optimal balance between CPU usage, memory allocation, and bandwidth utilization. Vieira et al. \cite{vieira.2022} propose a MILP solution to find the optimal allocation for tasks by maximizing the use of available computational resources and ensure an equitable distribution of services across servers that meet their Quality of Service (QoS) requirements. The proposed approach involves restricting the deployment of applications with lower sensitivity from being processed at the edge, instead redirecting them to the central cloud infrastructure. When an application arrives on the network, it initiates looking for an accessible server to handle its requests. This server can be either a MEC server or a cloud server. Additionally, it should be noted that the heuristic does not completely restrict the provisioning of applications on a particular node. Rather, it imposes a limitation on the number of applications of a specific kind that can be delivered on the MEC servers. In a large-scale scenario, the use of MILP becomes infeasible due to the high computational cost; hence, they propose a heuristic approach. Compared to GA proposed in \cite{maia.2021} and Particle Swarm Optimization proposed in \cite{geng.2021}, they enhance QoS requirements, such as latency and capacity of available resources, by optimizing the utilization of available resources \cite{vieira.2022}. 

In certain circumstances, simple tasks can be processed on mobile devices locally and the tasks that can't be processed locally can be offloaded to MEC. In special cases, it is possible to modify the CPU cycles allocated to Mobile devices/servers according to the complexity of the tasks \cite{li.2022}. A standard optimization approach has been proposed in \cite{yang.2021} to iteratively optimize the offloading data ratios and CPU frequency allocations. However, it is important to note that the algorithm does not take into account the selection of MEC resources. Following that, the actor-critic RL technique is employed to formulate a two-level learning-based algorithm to consider server selection and minimize energy consumption. The integration of a learning-based approach with optimization techniques at MEC servers is ultimately employed to facilitate the allocation of CPU frequencies at each MEC server \cite{yang.2021}. 

Through our comprehensive review of the existing literature in , we have identified a notable gap in current research on integrating the dropped task ratio with other network metrics. This article addresses the existing gap by focusing on the two primary goals of minimizing delay and dropped task ratio. In this study, we utilize two optimization approaches, namely MILP and GA for multi-user task offloading scenarios. The study commences with an analysis of the impact of networking factors, namely the number of users and tasks, on the metrics. In addition, we analyze the influence of these variables on scheduling algorithms, specifically FCFS and STF. 
\vspace{-1mm}
\section{System Model and Offloading Schemes}
We consider a multi-user 5G-based MEC, as shown in \figurename~\ref{fig:methodology}. The MEC server is deployed near a gNB to handle tasks offloaded from users and send back the results to them. We assume each user can have one or multiple tasks, and the server has two CPUs for processing them  as in \cite{truong.2020, hsu.2022,liao.2021}. We consider two different offloading schemes, FCFS and STF. Additionally, we aim to minimize delay and dropped task ratio by using MILP and Genetic Algorithm. 
\vspace{-3mm}
\subsection{ Greedy Schemes}
\textbf{FCFS} involves assigning tasks to computing resources based on their arrival order, without taking into account their computational demands or deadlines. Every incoming task is enqueued and awaits its turn for processing. The FCFS offloading approach, while straightforward to implement and equitable in terms of task execution order, can result in sub-optimal resource utilization and higher waiting times, particularly for smaller tasks that are queued behind larger, more time-consuming ones \cite{Ali2021}.

 \textbf{STF} is a scheduling approach that assigns priority to tasks according to their projected release time of the resources. In contrast to the FCFS scheduling algorithm, the STF algorithm prioritizes the execution of the shortest tasks first. The primary objective of this approach is to reduce the overall execution time and enhance the throughput of the system, and it aims for the task completion ratio. This approach improves the utilization of resources and has the potential to substantially decrease the average waiting time for tasks. Nevertheless, there is a potential drawback to this approach as it brings the possibility of longer tasks being held in queue, as they may persistently be assigned to lower priority positions \cite{Yin2022}.

\vspace{-5mm}
\begin{table}[!b]
\centering
\caption{Notation table for parameters used in the optimization problem.}
\begin{tabularx}{0.4\textwidth}{|c|X|}
\hline
\textbf{Parameter} & \textbf{Description} \\
\hline
\( N \) & Set of tasks \\
\hline
\( M \) & Set of CPUs \\
\hline
\( t_{w_i} \) & Waiting time of task \( i \) \\
\hline
\( W_i \) & Max of waiting time of task \( i \) \\
\hline
\( t_{d_i} \) & Deadline of task \( i \) \\
\hline
\( t_{a_i} \) & Arrival time of task \( i \) \\
\hline
\( t_{p_i} \) & Processing time of task \( i \) \\
\hline
\( t_{sp_i} \) & Start time of Processing task \( i \) \\
\hline
\( x_{ij} \) & Binary decision variable which is 1 when task i is assigned to CPU j.\\
\hline
\( M_{itj} \) & Binary decision variable which is 1 when task i is scheduled on CPU j in timeslot t.\\
\hline
\end{tabularx}

\label{tab:notation} \vspace{-2mm}
\end{table}

\subsection{The optimization problem}
The optimization problem includes an objective function that minimizes delay and dropped tasks. The objective function aims to make a trade-off between the benefits of the two baselines, FCFS and STF. The function integrates these two conflicting indicators into a cohesive purpose by utilizing a weighted sum method. The incorporation of a weighting factor, denoted as λ, provides the opportunity to adjust the optimization emphasis based on the particular needs of a given system. We also use  a binary decision variable, \( x_{ij} \), for assigning tasks to CPUs where:
\vspace{-1mm}
\begin{equation}
\hspace{-8mm}
  x_{ij} = 
  \begin{cases} 
    1 & \text{ task } i \text{ is assigned to CPU } j  \\
    0 & \text{otherwise}
  \end{cases}
  \label{eq:two}
\end{equation}

We formulate the optimization problem as Equation (\ref{eq:objective}), where \( N \) is the set of tasks,  \( M \) is the set of CPUs, \( t_{d_i} \) is the deadline of task \( i \), \( t_{a_i} \) is the arrival time of task \( i \),\( t_{w_i} \) is the waiting time of task \( i \), and \( t_{s_i} \) is the processing time of task \( i \). 
\vspace{-1mm}
\begin{align}
  \text{Min } \Bigg[ & \lambda \left( \sum_{i=1}^ {N} \sum_{j=1}^{M} x_{ij} \frac{t_{w_i}}{t_{d_i} - t_{a_i} - t_{s_i}} \right) \nonumber \\
  & + (1-\lambda) \left(\frac {\sum_{i=1}^{N} \sum_{j=1}^{M} (1 - x_{ij})}{N} \right) \Bigg]
  \label{eq:objective}
\end{align}

Start processing time constraint: Equation (\ref{eq:pt}) ensures that the processing of a a task starts after its arrival and there is enough time to complete the task before its deadline.
\begin{equation}
    t_{a_i}\leq t_{sp_i}  \leq t_{d_i} - t_{p_i}
    \label{eq:pt}
\end{equation}

Waiting time constraint: the waiting time for task i is non-negative and does not exceed the time available for execution before the deadline:
\begin{equation}
 0 \leq t_{w_i} \leq t_{d_i} - t_{a_i} - t_{p_i} \\
\end{equation}
 
Linearization constraint: We consider \( x_{ij} \cdot t_{w_i} \) as \( A_{ij}\) which can be linearized according to  (\ref{eq:three}) to (\ref{eq:five}), where \(W_i\) is maximum of \(t_{w_i}\) which is \(t_{d_i} - t_{a_i} - t_{p_i} \) :
\vspace{+1mm}
\begin{align}
A_{ij} - x_{ij} &\leq 0 \label{eq:three} \\
A_{ij} - t_{w_i} &\leq 0 \label{eq:four} \\
x_{ij} + t_{w_i} - A_{ij} &\leq W_i\label{eq:five}
\end{align}

CPU processing constraint: each task can be assigned to at most one CPU:
\vspace{-2mm}
\begin{equation}
\sum_{j=1}^ {M} x_{ij} \leq 1  \quad \forall i=1,...,N
\end{equation}

Scheduling constraint:  We define \(M_{itj}\) as a binary scheduling matrix for task i on CPU j in timeslot t, where \(M_{itj}=1\) if task i is scheduled on CPU j at timeslot t. Equation (\ref{eq:six}) ensures that no task starts before its arrival time. (\ref{eq:eight}) ensures that the total time spent on task \(i\) when scheduled on processor j during time slot t does not exceed the processing time \(t_{p_i}\):
\vspace{-1mm}
\begin{equation}
M_{itj} \geq t_{a_i} \quad \forall i,t,j
\label{eq:six}
\end{equation}

\begin{equation}
  \sum_{j} \sum_{t} x_{ij} \cdot M_{itj} \leq t_{p_i} \quad \forall i=1,...,N
  \label{eq:eight}
 \end{equation}
 
Where \(T_{itj} = x_{ij} \cdot M_{itj}\) will be linearized as  shown in (\ref{eq:nine}) to (\ref{eq:eleven}). Here, \(B_{itj}\) is the maximum value of \(M_{itj}\), which is 1.
\begin{align}
T_{itj} - x_{ij} &\leq 0 \label{eq:nine} \\
T_{itj} - M_{itj} &\leq 0 \label{eq:ten} \\
x_{ij} + M_{itj} - T_{itj} &\leq B_{itj}\label{eq:eleven}
\end{align}
\section{Performance Analysis}

The experiments were carried out using a computer equipped with a core i7 CPU and 32G RAM. The scenario has been simulated in ns3 and the parameters have been set according to \cite{Gao2023, Seah2022} as \tablename~\ref{tab:parameters}. The number of users and tasks have been chosen carefully to build a     realistic scenario. We set up the 5G gNB and connect it to MEC server. In MEC, there are two CPUs for processing tasks. We use a Poisson traffic generator to simulate random task arrival patterns to the edge servers. The tasks are distracted driver recognition by YOLO5. Tasks are offloaded with respect to the FCFS and STF order independently. Then we try to find the best offloading decision to minimize the dropped task ratio and delay. We choose different images in each run, but we use the same group of images in each run under different algorithms to have a fair comparison. The MILP and GA are used separately to minimize the objective function. The parameters used for GA are shown in \tablename~\ref{tab:GA}. These parameters have been chosen empirically after couple of experiments.The cxTwoPoint crossover and selTournament selection mechanism with size 3 in DEAP library have been used. The GA simulates 100 generations, each with 100 individuals. A 1\% mutation rate is applied to each gene. The mutation type is bit-string. The fitness function is defined as objective function in (\ref{eq:objective}).
 
\begin{table}[!hbt]
\centering
\caption{parameters values in simulation} 
\begin{tabular}{|c|c|}
\hline
\textbf{Parameters} & \textbf{Value}   \\   \hline
Number of Runs & 10   \\   \hline
Number of users & 10,100,500,1000   \\   \hline
Number of tasks per user& 1,5,10   \\   \hline
Number of CPU & 2   \\   \hline
CPU Speed &  2.2 Ghz   \\   \hline
Packet size & 1Kb   \\   \hline
Datarate & 50Mbps   \\   \hline

\end{tabular}
\label{tab:parameters} \vspace{-1mm}
\end{table}
\vspace{-3mm}
\begin{table}[h]
\centering 
\caption{Parameters for Genetic Algorithm}
\label{table:ga_parameters}
\begin{tabular}{|l|l|}
\hline
\textbf{Parameter} & \textbf{Value} \\
\hline
Population Size & 100 individuals in each generation \\
\hline
Number of Generations & 100 \\
\hline
Mutation Rate & 0.01 \\
\hline
\end{tabular}
\label{tab:GA}
\end{table}

\begin{figure}[!hbt]
 \centering
    \begin{subfigure}{0.5\textwidth} \vspace{-4mm}
        \centering
        \includegraphics[width = 0.65\textwidth, trim=1cm 0.0cm 1cm 1cm,clip]{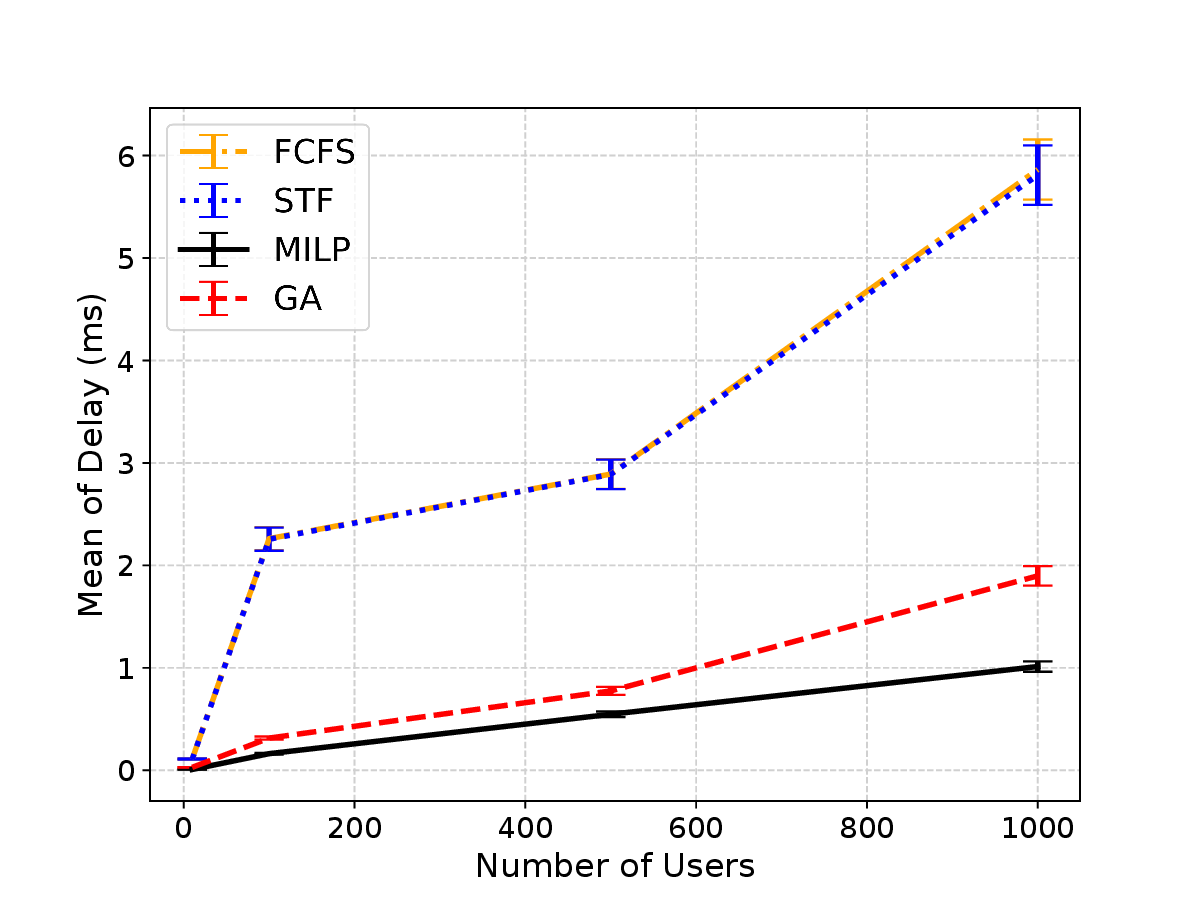}
        \caption{One task per user}
        \label{fig:d1t }
    \end{subfigure}
    \begin{subfigure}{0.5\textwidth}
        \centering
        \includegraphics[width = 0.65\textwidth, trim=1cm 0.0cm 1cm 1cm, clip]{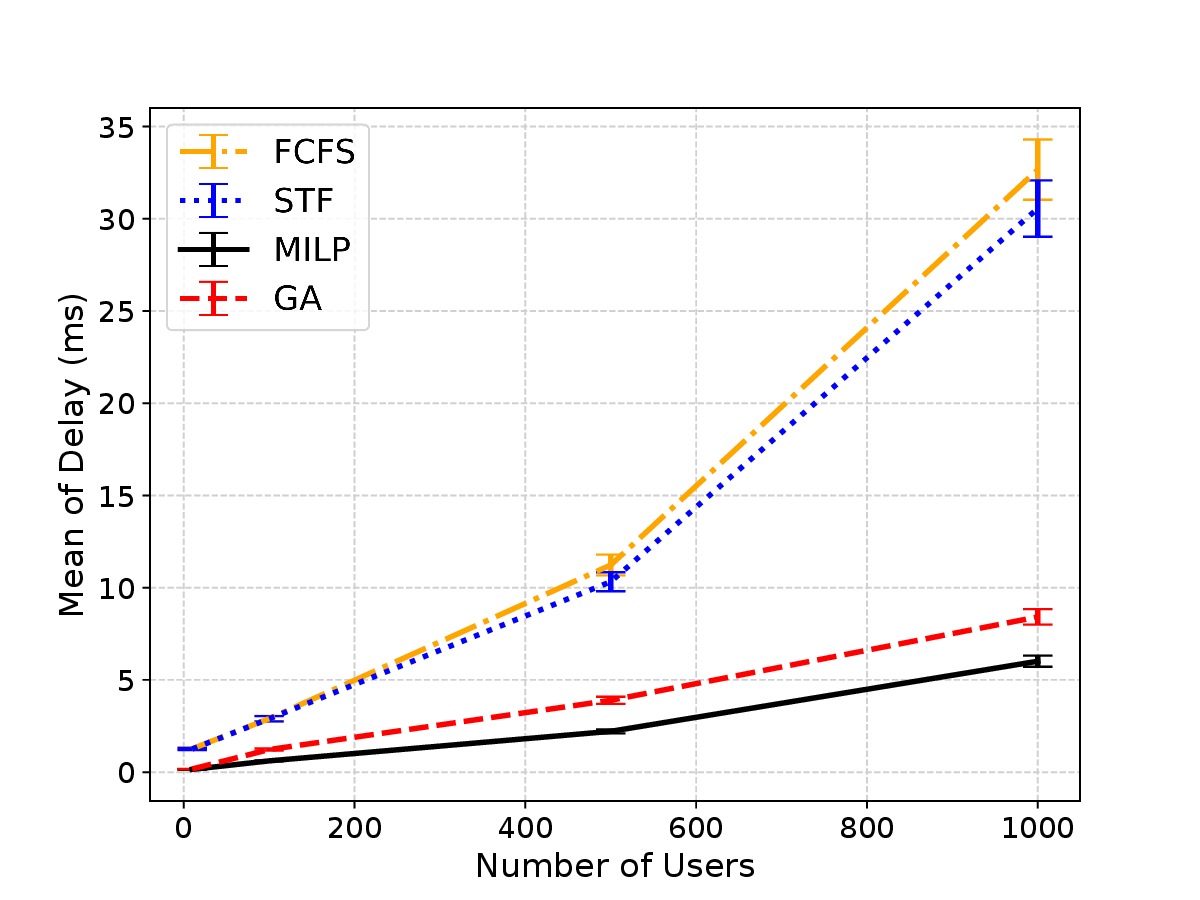}
        \caption{Five tasks per user}
        \label{fig:d5t}
    \end{subfigure}
    \begin{subfigure}{0.5\textwidth}
        \centering
        \includegraphics[width = 0.65\textwidth, trim=1cm 0.0cm 1cm 1cm, clip]{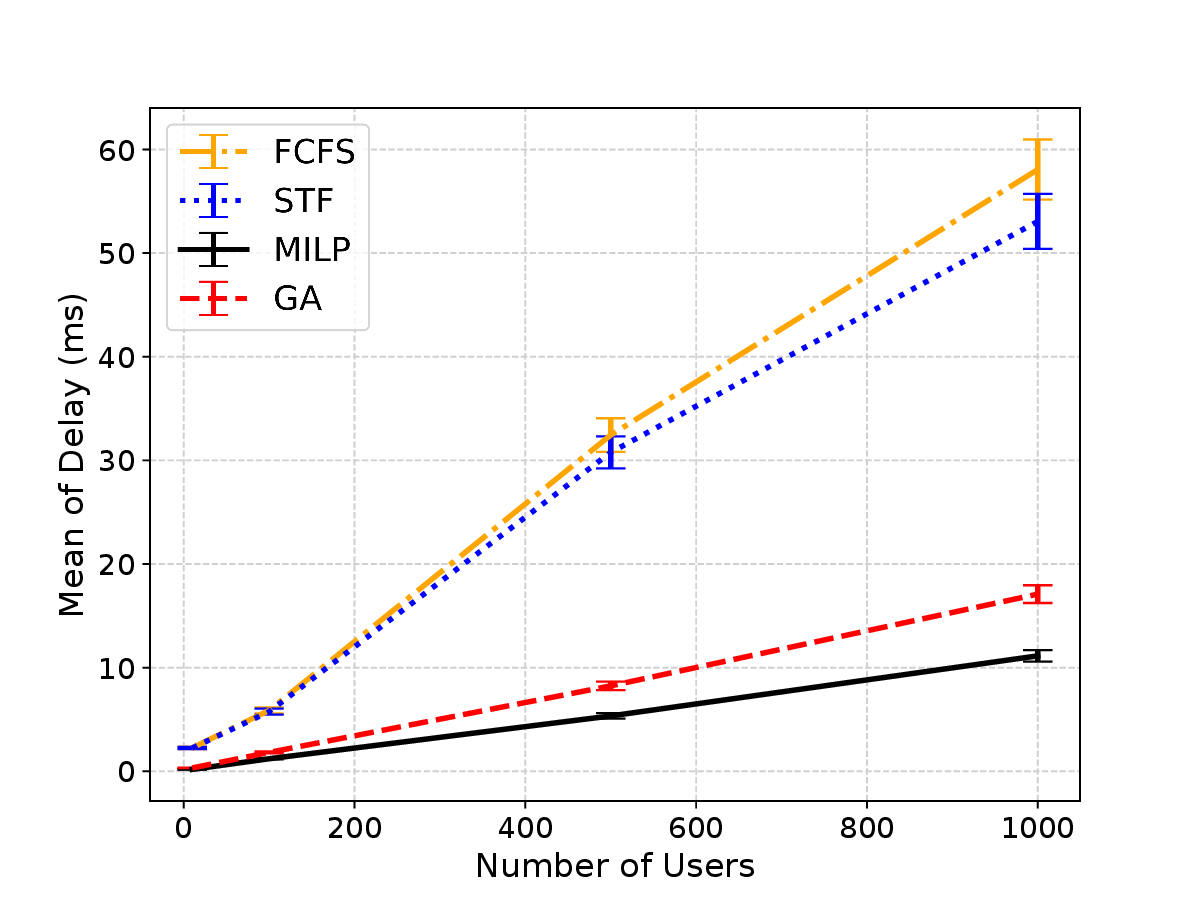}
        \caption{Ten tasks per user}
        \label{fig:d10t}
    \end{subfigure}
        \caption{Mean of delay for different number of tasks and confidence intervals}
	\label{fig:mean_delay} \vspace{-4mm}
\end{figure}

We plot the mean of delay for different numbers of tasks per user. As shown in \figurename~\ref{fig:mean_delay}, FCFS has the highest delay among all cases. When a long task is received just before too many short tasks, the shorter tasks will experience a delay in their execution as they wait  for the long task to be completed, resulting in an increase in their waiting time. 
In contrast, the STF scheduling algorithm assigns more priority to shorter tasks. In the context of task scheduling, it is observed that the arrival of a long task does not impede the execution of shorter tasks, resulting in a decrease in overall waiting time and subsequent reduction in delay. 

\begin{figure}[!t]
 \centering 
    \begin{subfigure}{0.5\textwidth}
        \centering
    \includegraphics[width = 0.7\textwidth, trim=0cm 0.3cm 1cm 1.83cm,clip]{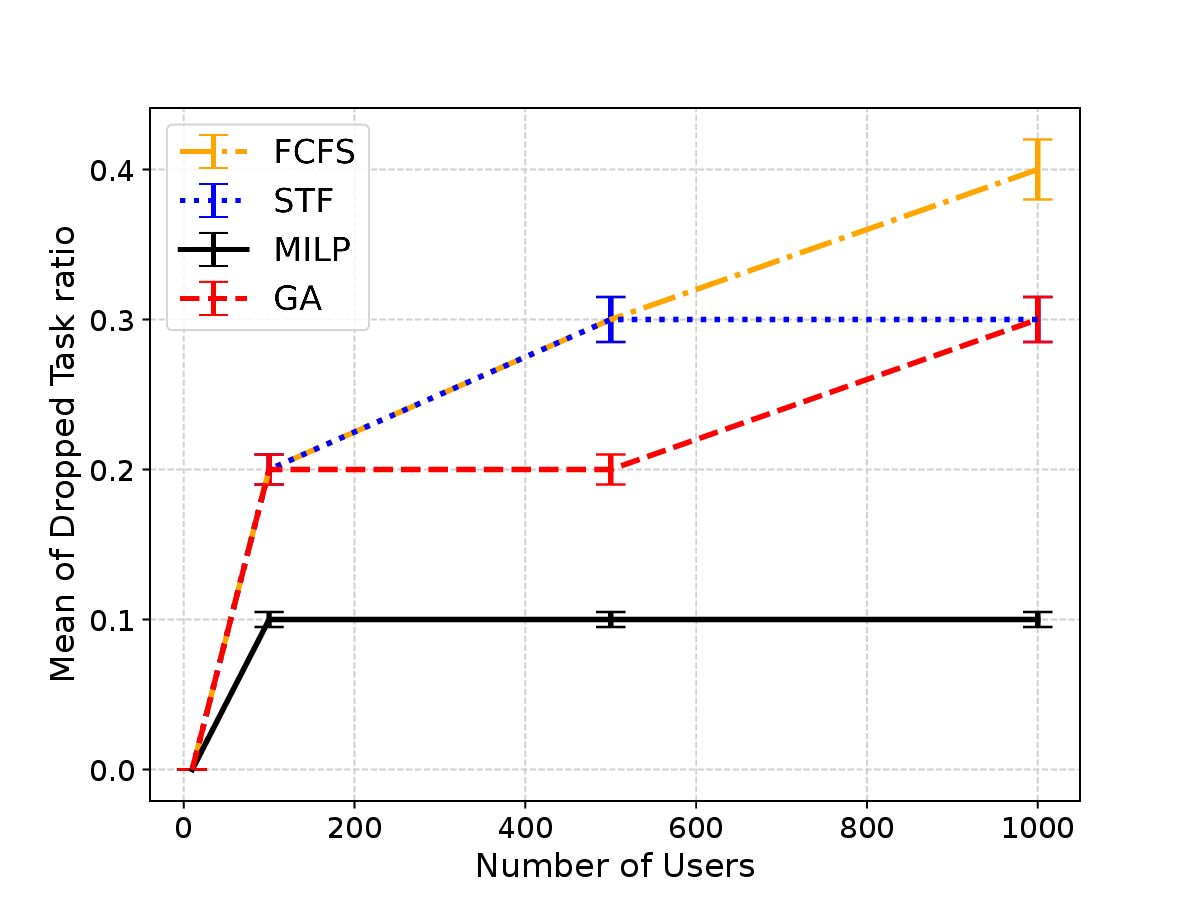}
        \caption{One task per user}
        \label{fig:mean_dtr1 }
    \end{subfigure}
    \begin{subfigure}{0.5\textwidth}
        \centering
        \includegraphics[width = 0.7\textwidth, trim=0cm 0.4cm 1cm 1.3cm, clip]{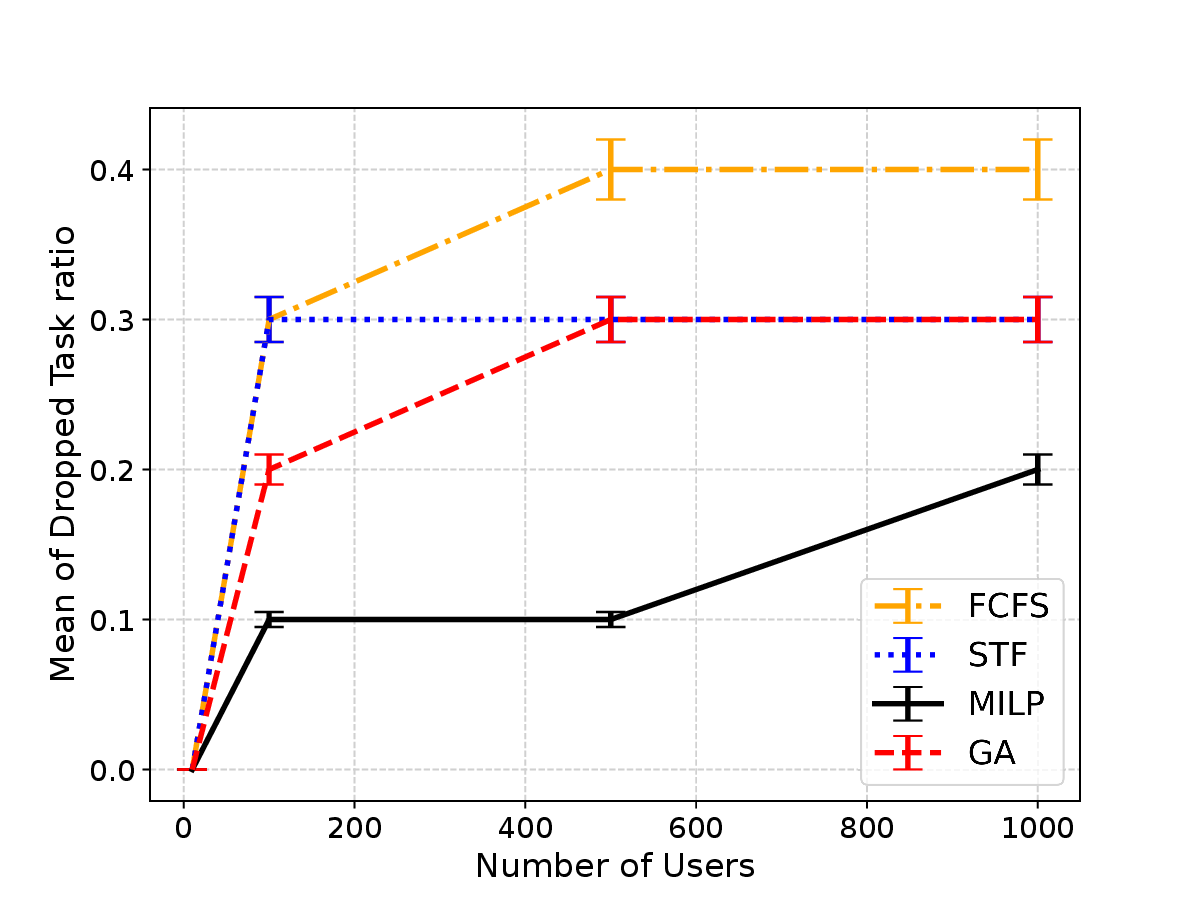}
        \caption{Five tasks per user}
        \label{fig:mean_dtr5}
    \end{subfigure}
    \begin{subfigure}{0.5\textwidth}
        \centering
        \includegraphics[width = 0.7\textwidth, trim=0cm 0.4cm 1cm 1.3cm, clip]{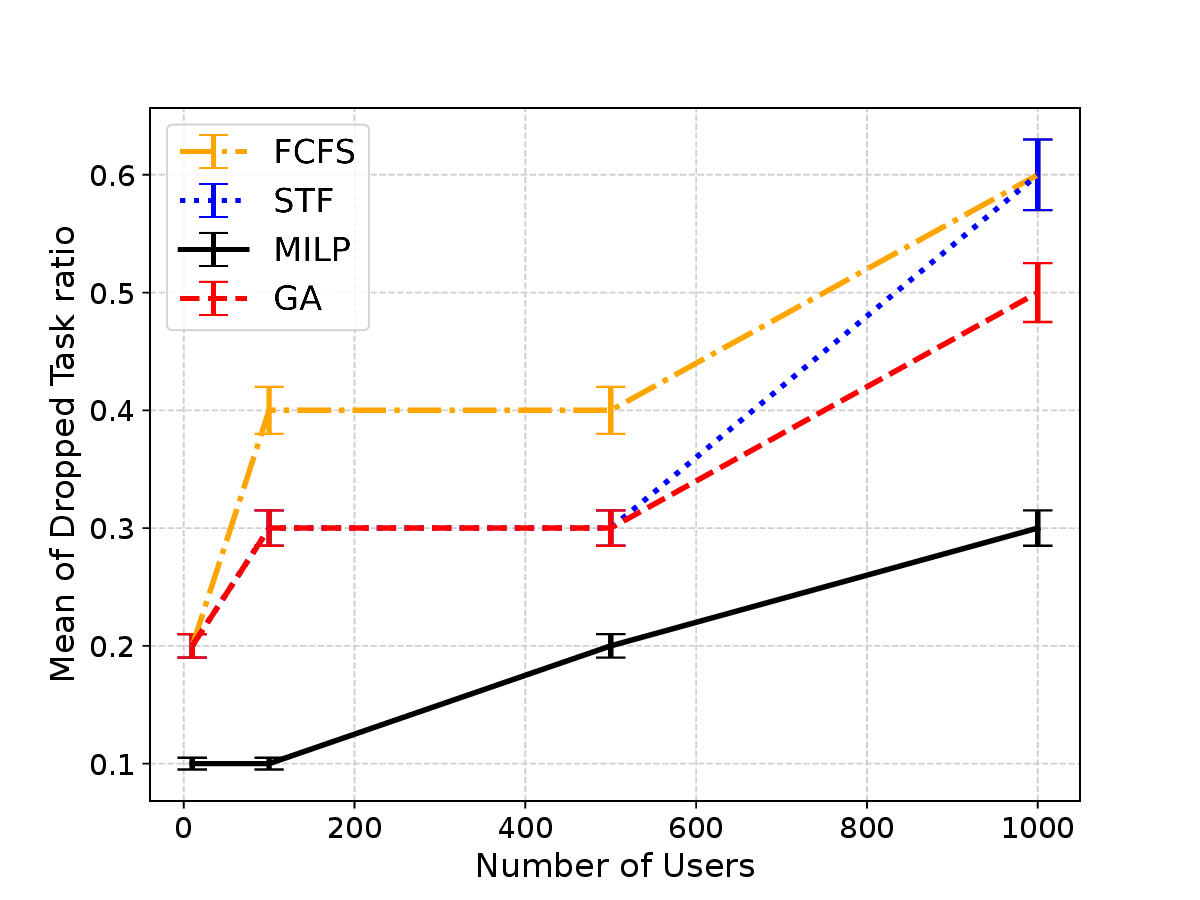}
        \caption{Ten tasks per user}
        \label{fig:mean_dtr10}
    \end{subfigure}
        \caption{Mean of dropped task ratio for different number of tasks and confidence intervals}
	\label{fig:mean_dtr} \vspace{-3mm}
\end{figure}

We also observe that MILP is slightly better than GA in terms of delay. It utilizes deterministic algorithms, which can yield an accurate answer given an adequate amount of time. When encountering problems that can be accurately represented by linear constraints and fall within the efficient scale of MILP, MILP has the capability to  identify the optimal solution faster. On the other hand, GA can be described as a meta-heuristic that operates by iteratively enhancing a population of solutions during a period of time. The iterative and evolutionary nature of the process in GA  might impose delays, particularly during the initial iterations. Therefore, in certain problem scenarios when both strategies produce almost similar results, MILP can attain more efficient results. In \tablename~\ref{tab:runtime_comparison}, MILP consistently outperforms GA in terms of run time. This indicates that for time-sensitive scenarios, MILP is the more optimal choice over GA.

\begin{table}[!b]
\centering\vspace{-2mm}
\caption{Run time comparison between MILP and GA}
\begin{tabular}{|c|c|c|} 
\hline
\multirow{2}{*}{Number of tasks} & \multicolumn{2}{c|}{Run time} \\
\cline{2-3}
& MILP & GA \\
\hline
10 & 42.2 ms & 62.5 ms \\
\hline
50 & 90.8 ms & 1.8 s \\
\hline
100 & 3.8 s & 4.6 s \\
\hline
500 & 11.6 s & 19.5 s \\
\hline
1000 & 21.8 s & 30.5 s \\
\hline
2500 & 40.3 s & 57.6 s \\
\hline
5000 & 60.1 s & 90.9 s \\
\hline
10000 & 125.6 s & 165.8 s \\
\hline
\end{tabular} 
\label{tab:runtime_comparison} 
\end{table}

We analyze the dropped task ratio for different cases in \figurename~\ref{fig:mean_dtr}. As shown in  \figurename~\ref{fig:mean_dtr}, MILP has the lowest dropped task ratio. If a system can be represented by linear equations, same as our optimization formula, the MILP can provide a solution that guarantees the lowest feasible dropped task ratio. The GA requires the iterative evolution of a population of solutions. The efficiency of GA in minimizing the dropped task ratio depends on the problem representation. The system performance is significantly influenced by its tuning.

The STF approach  can increase the number of tasks processed within a specific time period, potentially reducing the ratio of dropped tasks. This is because shorter tasks are eliminated more quickly, resulting in the freeing up of resources. In the FCFS  scheduling algorithm, when there is a shortage of resources and a long task arrives before  multiple shorter tasks, the  queue length increases significantly, resulting in a potential rise in the ratio of dropped tasks, particularly during periods of high demand. Tables 2 to 4  provide detailed values of delay and dropped task ratio for different cases. Overall,as seen in \figurename~\ref{fig:mean_delay} and \figurename~\ref{fig:mean_dtr}, MILP results in shorter delays and dropped task ratio than other methods, making it a viable solution for addressing our problem. Additionally, according to \tablename~\ref{tab:runtime_comparison}, MILP run time is significantly shorter than GA, making it the optimal solution for our case.

\vspace{-2mm}
\section{Conclusion}

 An approach in task offloading has been proposed, considering computing delay and dropped task ratio, offering a holistic view of system efficiency in a realistic and scalable 5G-MEC scenario. The real-time video frame (image) analysis by YOLO5 is considered as a computationally heavy task that will be offloaded to an edge server. By employing MILP and GA, our optimization techniques have shown to outperform the baseline heuristics, such as FCFS and STF. In MILP, the dropped task ratio and delay has been minimized by 20\% and 2ms compared to GA. Our results prove to be highly effective for optimizing service performance within the 5G-MEC context. Analytical model will be extended with more elements of real use cases in the future. More MEC servers will be incorporated into our simulations. Moreover, several schedule optimization strategies will be investigated to improve the systems overall efficiency by effectively utilizing available resources and addressing difficulties in coping with increased demand. 
 
\vspace{-2mm}
\section*{Acknowledgment}
\vspace{-0.05in}

This work was supported in part by funding from the Innovation for Defence Excellence and Security (IDEaS) program from the Department of National Defence (DND) and in part by NSERC CREATE TRAVERSAL Program.
\vspace{-3mm}
\bibliographystyle{IEEEtran}

\begin{thebibliography}{10}
\providecommand{\url}[1]{#1}
\csname url@samestyle\endcsname
\providecommand{\newblock}{\relax}
\providecommand{\bibinfo}[2]{#2}
\providecommand{\BIBentrySTDinterwordspacing}{\spaceskip=0pt\relax}
\providecommand{\BIBentryALTinterwordstretchfactor}{4}
\providecommand{\BIBentryALTinterwordspacing}{\spaceskip=\fontdimen2\font plus
\BIBentryALTinterwordstretchfactor\fontdimen3\font minus
  \fontdimen4\font\relax}
\providecommand{\BIBforeignlanguage}[2]{{%
\expandafter\ifx\csname l@#1\endcsname\relax
\typeout{** WARNING: IEEEtran.bst: No hyphenation pattern has been}%
\typeout{** loaded for the language `#1'. Using the pattern for}%
\typeout{** the default language instead.}%
\else
\language=\csname l@#1\endcsname
\fi
#2}}
\providecommand{\BIBdecl}{\relax}
\BIBdecl

\bibitem{haibeh.2022}
L.~A. Haibeh, M.~C.~E. Yagoub, and A.~Jarray, ``A survey on mobile edge
  computing infrastructure: Design, resource management, and optimization
  approaches,'' \emph{IEEE Access}, vol.~10, pp. 27\,591--27\,610, 2022.

\bibitem{kumaran.2021}
K.~Kumaran and E.~Sasikala, ``Learning based latency minimization techniques in
  mobile edge computing (mec) systems: A comprehensive survey,'' in \emph{Intl
  Conf. Sys., Comp, Automation and Netw.}, 2021, pp. 1--6.

\bibitem{nencioni2023}
G.~Nencioni, R.~G. Garroppo, and R.~F. Olimid, ``5g multi-access edge
  computing: a survey on security, dependability, and performance,'' \emph{IEEE
  Access}, 2023.

\bibitem{Cruz2022}
\BIBentryALTinterwordspacing
P.~Cruz, N.~Achir, and A.~C. Viana, ``On the edge of the deployment: A survey
  on multi-access edge computing,'' \emph{ACM Comput. Surv.}, vol.~55, no.~5,
  dec 2022. [Online]. Available: \url{https://doi.org/10.1145/3529758}
\BIBentrySTDinterwordspacing

\bibitem{hua2023edge}
H.~Hua, Y.~Li, T.~Wang, N.~Dong, W.~Li, and J.~Cao, ``Edge computing with
  artificial intelligence: A machine learning perspective,'' \emph{ACM
  Computing Surveys}, vol.~55, no.~9, pp. 1--35, 2023.

\bibitem{sharma.2022}
A.~Sharma, C.~Diwaker, and M.~Nadiyan, ``Analysis of offloading computation in
  mobile edge computing (mec): A survey,'' in \emph{2022 Seventh International
  Conference on Parallel, Distributed and Grid Computing (PDGC)}, 2022, pp.
  280--285.

\bibitem{feng.2021}
S.~Feng, Y.~Chen, Q.~Zhai, M.~Huang, and F.~Shu, ``Optimizing computation
  offloading strategy in mobile edge computing based on swarm intelligence
  algorithms,'' \emph{EURASIP Journal on Advances in Signal Processing}, vol.
  2021, no.~1, pp. 1--15, 2021.

\bibitem{truong.2020}
T.~P. Truong, A.-T. Tran, A.~Masood, D.~S. Lakew, C.~Lee, Y.~Lee, S.~Cho
  \emph{et~al.}, ``Delay-sensitive task offloading for internet of things in
  nonorthogonal multiple access mec networks,'' in \emph{2020 International
  Conference on Information and Communication Technology Convergence
  (ICTC)}.\hskip 1em plus 0.5em minus 0.4em\relax IEEE, 2020, pp. 597--599.

\bibitem{zhu.2021}
A.~Zhu and Y.~Wen, ``Computing offloading strategy using improved genetic
  algorithm in mobile edge computing system,'' \emph{Journal of Grid
  Computing}, vol.~19, no.~3, p.~38, 2021.

\bibitem{liao.2021}
Z.~Liao, J.~Peng, B.~Xiong, and J.~Huang, ``Adaptive offloading in mobile-edge
  computing for ultra-dense cellular networks based on genetic algorithm,''
  \emph{Journal of Cloud Computing}, vol.~10, no.~1, pp. 1--16, 2021.

\bibitem{hsu.2022}
C.-K. Hsu, ``A dueling dqn-based computational offloading method in mec-enabled
  iiot network,'' \emph{The Computer Journal}, p. bxac133, 2022.

\bibitem{liu.2019}
M.~Liu, F.~R. Yu, Y.~Teng, V.~C. Leung, and M.~Song, ``Performance optimization
  for blockchain-enabled industrial internet of things (iiot) systems: A deep
  reinforcement learning approach,'' \emph{IEEE Trans. on Industrial
  Informatics}, vol.~15, no.~6, pp. 3559--3570, 2019.

\bibitem{wang.2019}
J.~Wang, C.~Jiang, K.~Zhang, X.~Hou, Y.~Ren, and Y.~Qian, ``Distributed
  q-learning aided heterogeneous network association for energy-efficient
  iiot,'' \emph{IEEE Transactions on Industrial Informatics}, vol.~16, no.~4,
  pp. 2756--2764, 2019.

\bibitem{yuan.2022}
X.~Yuan, H.~Tian, Z.~Zhang, Z.~Zhao, L.~Liu, A.~K. Sangaiah, and K.~Yu, ``A mec
  offloading strategy based on improved dqn and simulated annealing for
  internet of behavior,'' \emph{ACM Transactions on Sensor Networks}, vol.~19,
  no.~2, pp. 1--20, 2022.

\bibitem{vieira.2022}
R.~F. Vieira, D.~D.~S. Souza, M.~S. Da~Silva, and D.~L. Cardoso, ``A heuristic
  for load distribution on data center hierarchy: A mec approach,'' \emph{IEEE
  Access}, vol.~10, pp. 69\,462--69\,471, 2022.

\bibitem{maia.2021}
A.~M. Maia, Y.~Ghamri-Doudane, D.~Vieira, and M.~F. de~Castro, ``An improved
  multi-objective genetic algorithm with heuristic initialization for service
  placement and load distribution in edge computing,'' \emph{Computer
  networks}, vol. 194, p. 108146, 2021.

\bibitem{geng.2021}
N.~Geng, Z.~Chen, Q.~A. Nguyen, and D.~Gong, ``Particle swarm optimization
  algorithm for the optimization of rescue task allocation with uncertain time
  constraints,'' \emph{Complex \& Intelligent Systems}, vol.~7, pp. 873--890,
  2021.

\bibitem{li.2022}
H.~Li, K.~D.~R. Assis, S.~Yan, and D.~Simeonidou, ``Drl-based long-term
  resource planning for task offloading policies in multiserver edge computing
  networks,'' \emph{IEEE Transactions on Network and Service Management},
  vol.~19, no.~4, pp. 4151--4164, 2022.

\bibitem{yang.2021}
Y.~Yang and M.~C. Gursoy, ``Optimization and learning for data offloading and
  resource management in mobile edge computing,'' in \emph{2021 IEEE
  International Conference on Pervasive Computing and Communications Workshops
  and other Affiliated Events (PerCom Workshops)}.\hskip 1em plus 0.5em minus
  0.4em\relax IEEE, 2021, pp. 598--603.

\bibitem{Ali2021}
\BIBentryALTinterwordspacing
A.~Ali, M.~M. Iqbal, H.~Jamil, F.~Qayyum, S.~Jabbar, O.~Cheikhrouhou, M.~Baz,
  and F.~Jamil, ``An efficient dynamic-decision based task scheduler for task
  offloading optimization and energy management in mobile cloud computing,''
  \emph{Sensors}, vol.~21, no.~13, 2021. [Online]. Available:
  \url{https://www.mdpi.com/1424-8220/21/13/4527}
\BIBentrySTDinterwordspacing

\bibitem{Yin2022}
G.~Yin, R.~Chen, and Y.~Zhang, ``Effective task offloading heuristics for
  minimizing energy consumption in edge computing,'' in \emph{2022 IEEE
  International Conferences on Internet of Things (iThings) and IEEE Green
  Computing \& Communications (GreenCom) and IEEE Cyber, Physical \& Social
  Computing (CPSCom) and IEEE Smart Data (SmartData) and IEEE Congress on
  Cybermatics (Cybermatics)}, 2022, pp. 243--249.

\bibitem{Gao2023}
M.~Gao, R.~Shen, L.~Shi, W.~Qi, J.~Li, and Y.~Li, ``Task partitioning and
  offloading in dnn-task enabled mobile edge computing networks,'' \emph{IEEE
  Transactions on Mobile Computing}, vol.~22, no.~4, pp. 2435--2445, 2023.

\bibitem{Seah2022}
W.~K. Seah, C.-H. Lee, Y.-D. Lin, and Y.-C. Lai, ``Combined communication and
  computing resource scheduling in sliced 5g multi-access edge computing
  systems,'' \emph{IEEE Transactions on Vehicular Technology}, vol.~71, no.~3,
  pp. 3144--3154, 2022.

\end{thebibliography}

\end{document}